\documentclass[twocolumn]{aastex62}
\usepackage{subfigure}
\usepackage{amsmath}    
\usepackage{amssymb}    
\usepackage{bm}
\usepackage{url}
\usepackage{upgreek}
\usepackage{xspace}

\newcommand{\ppcc}{$\,$pc$\,$cm$^{-3}$} 
\newcommand{\mi}{\ensuremath{m_I}\xspace}
\newcommand{\misq}{\ensuremath{m_I^2}\xspace}
\newcommand{\ibar}{\ensuremath{\overline{I}}\xspace}
\newcommand{\ibarsq}{\ensuremath{\overline{I}^2}\xspace}
\newcommand{\isqbar}{\ensuremath{\overline{I^2}}\xspace}
\def\magn{J1810$-$197\xspace}
\def\mpsr{J1809$-$1943\xspace}
\def\rfrb{FRB~121102\xspace}
\def\tausc{$\tau_{\rm sc}$\xspace}
\def\tsky{$T_{\rm sky}$\xspace}

\shorttitle{Burst emission properties of \magn}
\shortauthors{Maan et al.}
\begin{document}
\title{Distinct properties of the radio burst emission from the magnetar XTE~\magn}
\correspondingauthor{Yogesh Maan}
\email{maan@astron.nl}
\author[0000-0002-0862-6062]{Yogesh Maan}
\affil{ASTRON, Netherlands Institute for Radio Astronomy, Oude Hoogeveensedijk 4, 7991 PD, Dwingeloo, The Netherlands}
\author[0000-0002-0863-7781]{Bhal Chandra Joshi}
\affiliation{National Centre for Radio Astrophysics, Tata Institute of Fundamental Research, Post Bag 3, Ganeshkhind, Pune 411007, India}
\author[0000-0002-9507-6985]{Mayuresh P. Surnis}
\affiliation{West Virginia University, Department of Physics and Astronomy, P.O. Box 6315, Morgantown, WV 26506, USA}
\affiliation{Center for Gravitational Waves and Cosmology, West Virginia University, Chestnut Ridge Research Building, Morgantown, WV 26505, USA}
\author[0000-0001-8640-8186]{Manjari Bagchi}
\affiliation{The Institute of Mathematical Sciences, C.I.T. Campus, Taramani, Chennai, 600113, India}
\affiliation{Homi Bhabha National Institute, Training School Complex, Anushakti Nagar, Mumbai 400094, India}
\author[0000-0003-4274-211X]{P. K. Manoharan}
\affiliation{National Centre for Radio Astrophysics, Tata Institute of Fundamental Research, Post Bag 3, Ganeshkhind, Pune 411007, India}
\affiliation{Radio Astronomy Centre, NCRA-TIFR, Udagamandalam, India}

\begin{abstract}
XTE~\magn (PSR J1809$-$1943) was the first ever magnetar which was found to emit transient radio
emission. It has recently undergone another radio and high-energy outburst. This
is only the second radio outburst that has been observed from this source.
We observed \magn soon after its recent radio outburst at low radio frequencies
using the Giant Metrewave Radio Telescope. We present the 650\,MHz flux density
evolution of the source in the early phases of the outburst, and its radio
spectrum down to frequencies as low as 300\,MHz. The magnetar also exhibits
radio emission in the form of strong, narrow bursts. We show that the bursts
have a characteristic intrinsic width of the order of 0.5$-$0.7\,ms, and discuss
their properties
in the context of giant pulses and giant micropulses from other pulsars.
We also show that the bursts exhibit spectral structures which cannot
be explained by interstellar propagation effects. These structures might
indicate a phenomenological link with the repeating fast radio bursts which
also show interesting, more detailed frequency structures. While the spectral
structures are particularly noticeable in the early phases of the outburst,
these seem to be less prominent as well as less frequent in the later phases,
suggesting an evolution of the underlying cause of these spectral structures. 
\end{abstract}


\section{Introduction} \label{sec-intro}
Magnetars are characterized by their high magnetic fields
($10^{14}-10^{15}$\,G), young age, persistent but highly variable X-ray
emission, and transient radio emission. Transient radio pulsations have
been observed from a handful of magnetars, while no radio emission has
been found from others despite deep searches \citep[e.g.,][]{Surnis16}.
The anomalous X-ray pulsar XTE~\magn (PSR~\mpsr) was the first magnetar
found to be emitting radio pulses \citep{Camilo06} after a strong high
energy outburst \citep{Gotthelf03,Ibrahim04}. At the beginning of the
outburst, a nearly flat spectral index between
0.7$-$42\,GHz was reported. The radio flux density decreased along with
the X-ray flux with a long decay time, and the source became
undetectable in late 2008 \citep{Camilo16}. Regular radio monitoring of the
source revealed its reactivation at 1.5\,GHz in late 2018 \citep{Lyne18},
which was followed by its successful detection at a wide range of radio
frequencies \citep[0.65$-$11.7\,GHz, e.g.,][]{Joshi18,Trushkin19}.
\par
In its previous outburst, \magn exhibited spikes or bursts of radio emission
with typical widths $\lesssim$10\,ms, and structures as narrow as 0.2\,ms
\citep{Camilo06}.
There are hints that the current outburst
also exhibits millisecond-width bright pulses \citep{Dai19}. A number of
similar emission components from other classes of pulsar population are known,
e.g., the giant-pulses \citep[][]{SR68,WCS84,Joshi04,MAD12,Maan15}, the giant
micropulses from the Vela pulsar \citep{Johnston01}, spiky emission from
PSR~B0656+14 \citep{Weltevrede06}, PSR~J0437$-$4715 \citep{Ables97,Vivekanand00},
and rotating radio transients \citep[][]{McLaughlin06}. Any similarity
between the spiky emission from the magnetar and the above mentioned emission
components could provide an important link between the corresponding emission
mechanisms. A study of the narrow, bright bursts from the magnetar could also
provide further clues to the origin of fast radio bursts (FRBs) ---
milliseconds-wide highly luminous radio transient events, most-likely of
extra-galactic origin \citep{Lorimer07,Thornton13}. There are indeed a number
of models which invoke magnetars \citep[e.g.,][]{MM18} as the sources of FRBs.
\par
To the best of our knowledge, there has been only one detailed single pulse
study of this object during its previous outburst. \citet{Serylak09} used
multi-frequency observations
to conduct detailed fluctuation analysis and study pulse-energy distributions,
and discussed the nature of the magnetar's single pulses in relevance to some
of the above mentioned emission components. However, we note that this study
was limited by a temporal resolution of 5\,ms.
\par
Here we present observations of the spiky or bursty emission from the magnetar
at a number of epochs during the current outburst. We characterize various
properties of the narrow bursts and discuss their relevance with the similar
emission components from pulsars and FRBs. We also present the 650\,MHz
flux density variations during first few months of the magnetar's recent
outburst, and the radio spectrum at the initial phases of the outburst down to
frequencies as low as 300\,MHz.
\par
\begin{deluxetable*}{lccccl}
\tablecaption{Details of observations and power-law fits to the
burst peak flux densities. \label{tab1}}
\tablecolumns{6}
\tablenum{1}
\tablewidth{0pt}
\tablehead{
  \colhead{Session~ID}       &
  \colhead{Date}             &
  \colhead{Frequency}           &
  \colhead{Sampling}         &
  \colhead{Duration}         &
  \colhead{Fitted power-law}       \\
  \colhead{}                 &
  \colhead{(YYYY-mm-dd)}     &
  \colhead{Band (MHz)}  &
  \colhead{Time (ms)}        &
  \colhead{(minutes)}   &
  \colhead{index ($\alpha$)}     
}
\startdata
S1       & 2018-12-18 & 550$-$750 (B4)  &  1.31072 & 40 & $-3.4\pm0.2$ \\
S2       & 2018-12-28 & 550$-$750 (B4)  &  0.16384 & 35 & $-3.8\pm0.2$ \\
S3       & 2019-02-15 & 550$-$750 (B4)  &  0.65536 & 28 & $-3.3\pm0.2$ \\
S4$_{a}$ & 2019-02-17 & 550$-$750 (B4)  &  0.65536 & 20 & $-3.0\pm0.3$ \\
S4$_{b}$ & 2019-02-17 & 1260$-$1460 (B5)&  0.65536 & 20 & $-1.95\pm0.3$ \\
\enddata
\tablecomments{The observations in sessions S4$_{a}$ and S4$_{b}$ were simultaneous.}
\end{deluxetable*}

\begin{figure}
\centering
\subfigure{\includegraphics[width=0.48\textwidth]{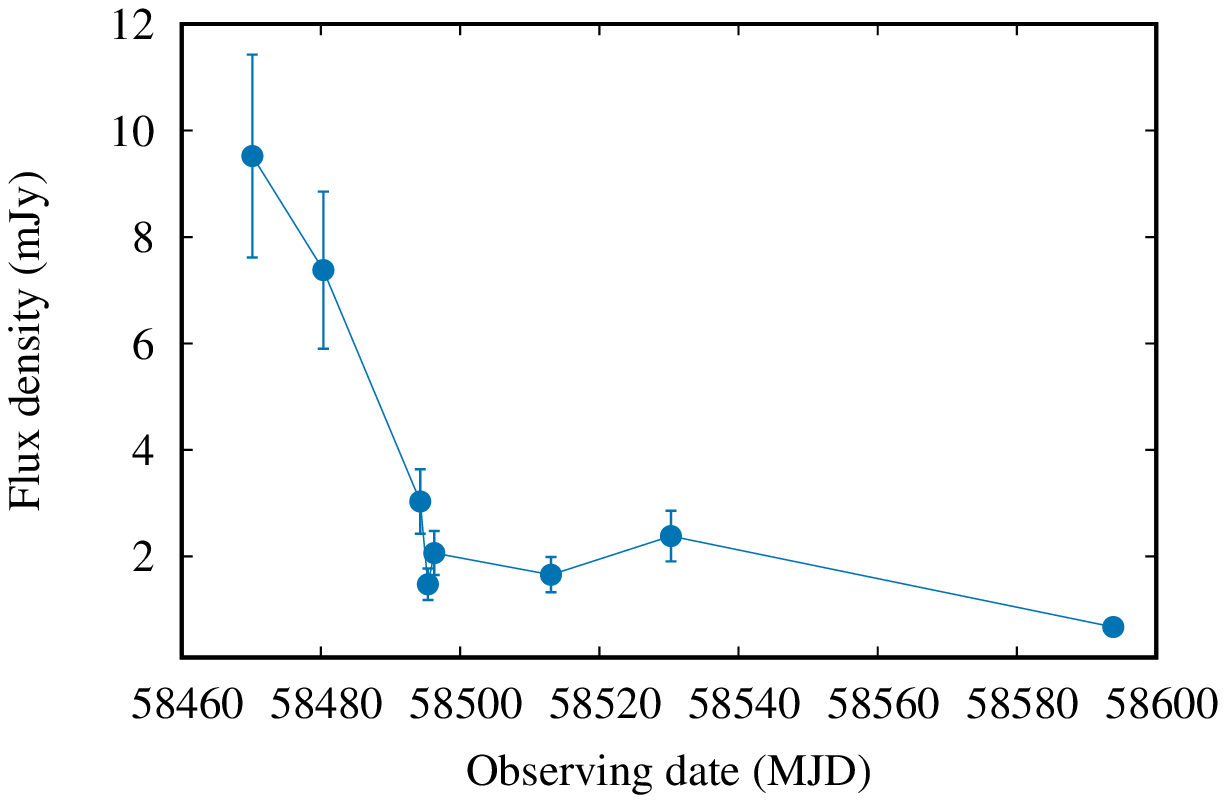}} 
\subfigure{\includegraphics[width=0.48\textwidth]{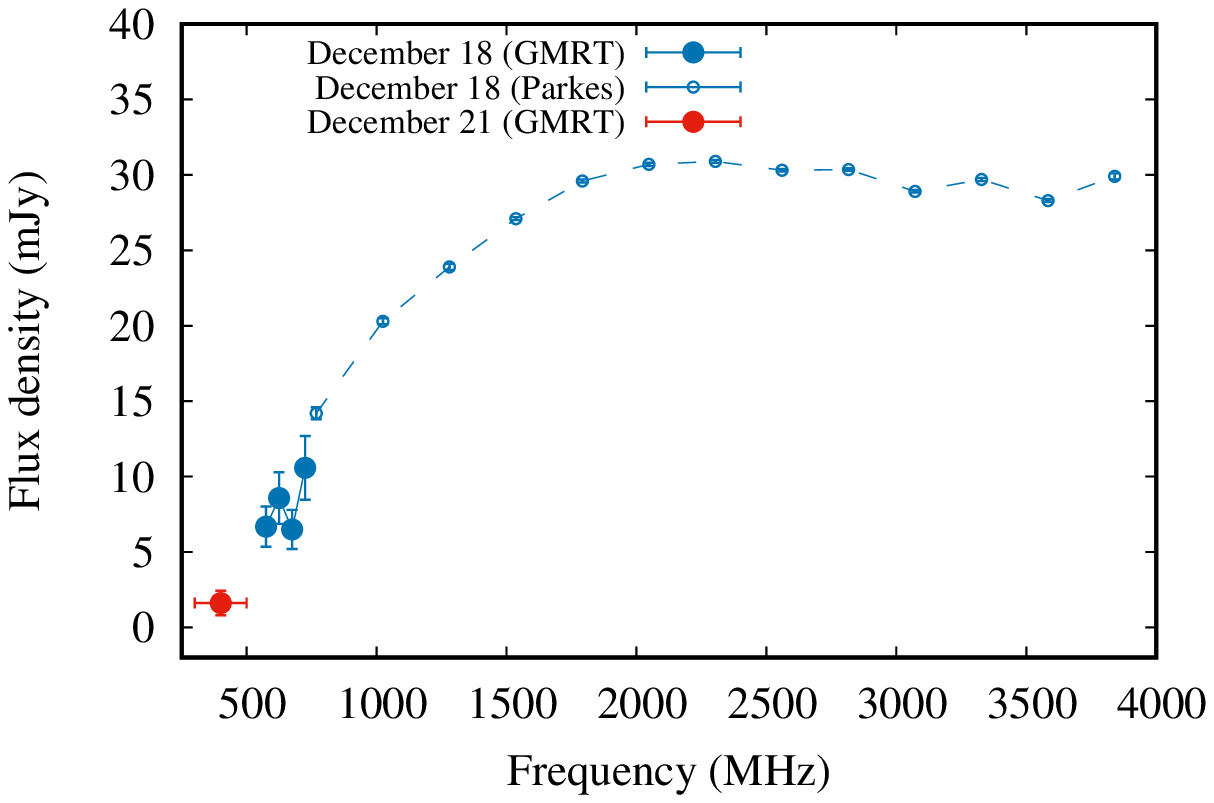}}
\caption{\textsl{Top panel:} The temporal evolution of the period-averaged
flux density of the magnetar in the frequency range 550$-$750\,MHz. The error bars
are arbitrarily assumed to be 20\% of the measurements. \textsl{Bottom panel:}
The flux density spectrum including our measurements at frequencies below
750\,MHz and those from \citet{Dai19} at higher frequencies.
\label{fig-flux}}
\end{figure}
\section{Observations and data reduction} \label{sec-obs}
\magn was observed at a number of epochs between
December 18, 2018 and February 17, 2019 with the upgrade giant metrewave radio
telescope \citep[uGMRT;][]{Gupta17} using the director's discretionary time allocations
(proposal~IDs: ddtC042 and ddtC044). These pulsar-mode observations
utilized the recently upgraded wide-band backends to record 200\,MHz
bandwidth.
In this letter, we present most of our results on the magnetar's spiky
emission obtained primarily from 4 observations in the 550$-$750\,MHz band
(see Table~\ref{tab1}). In one of these sessions, we combined the GMRT
antennae in two sub-arrays to also simultaneously observe in the frequency
band 1260$-$1460\,MHz. The flux density evolution presented in
Section~\ref{subsec-flux} includes measurements from observations not
discussed otherwise, including a recent April 2019 observation from our
long-term monitoring campaign of this source (proposal~ID 36\_082).
\par
For each of the epochs, we used the pulsar search and analysis software
\textsl{PRESTO} \citep{RansomThesis} to excise radio frequency interference,
and compute time sequences dedispersed to a dispersion measure of
178.5\,\ppcc. We used these time sequences for the results presented in
the following section.
\begin{figure*}
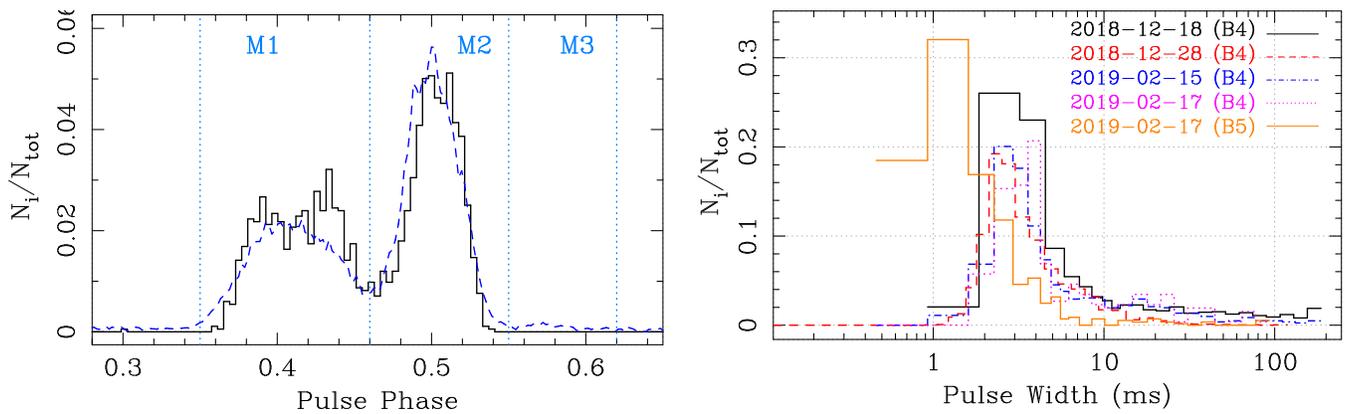

\centering
\subfigure{\includegraphics[width=0.295\textwidth,angle=-90]{b4-20181218sp_sw_sphist.ps}}
\hspace*{2mm}
\subfigure{\includegraphics[width=0.295\textwidth,angle=-90]{all_combined_whist.ps}} 
\caption{\textsl{Left:} The black curve shows the histogram of bright pulses
as a function of the spin phase, while the dashed, blue curve shows the
average profile shape from session~S1. The vertical light blue colored dotted
lines mark the assumed extents of the individual components (used in
Figure~\ref{fig-peakhist} to examine the component-separated peak flux
density distributions).
\textsl{Right:} Pulse-width distributions of bursts in
several observing sessions are shown. The legends indicate the date of
observing session as well as the observing band (B4:~550$-$750\,MHz and
B5:~1260$-$1460\,MHz). It is clearly evident that the 550$-$750\,MHz
observations exhibit a characteristic \emph{scatter-broadened} pulse-width
of a few ms, while it is much narrower at the higher frequencies.
\label{fig-sp}}
\end{figure*}
\begin{figure*}
\centering
\includegraphics[width=0.75\textwidth,angle=-90]{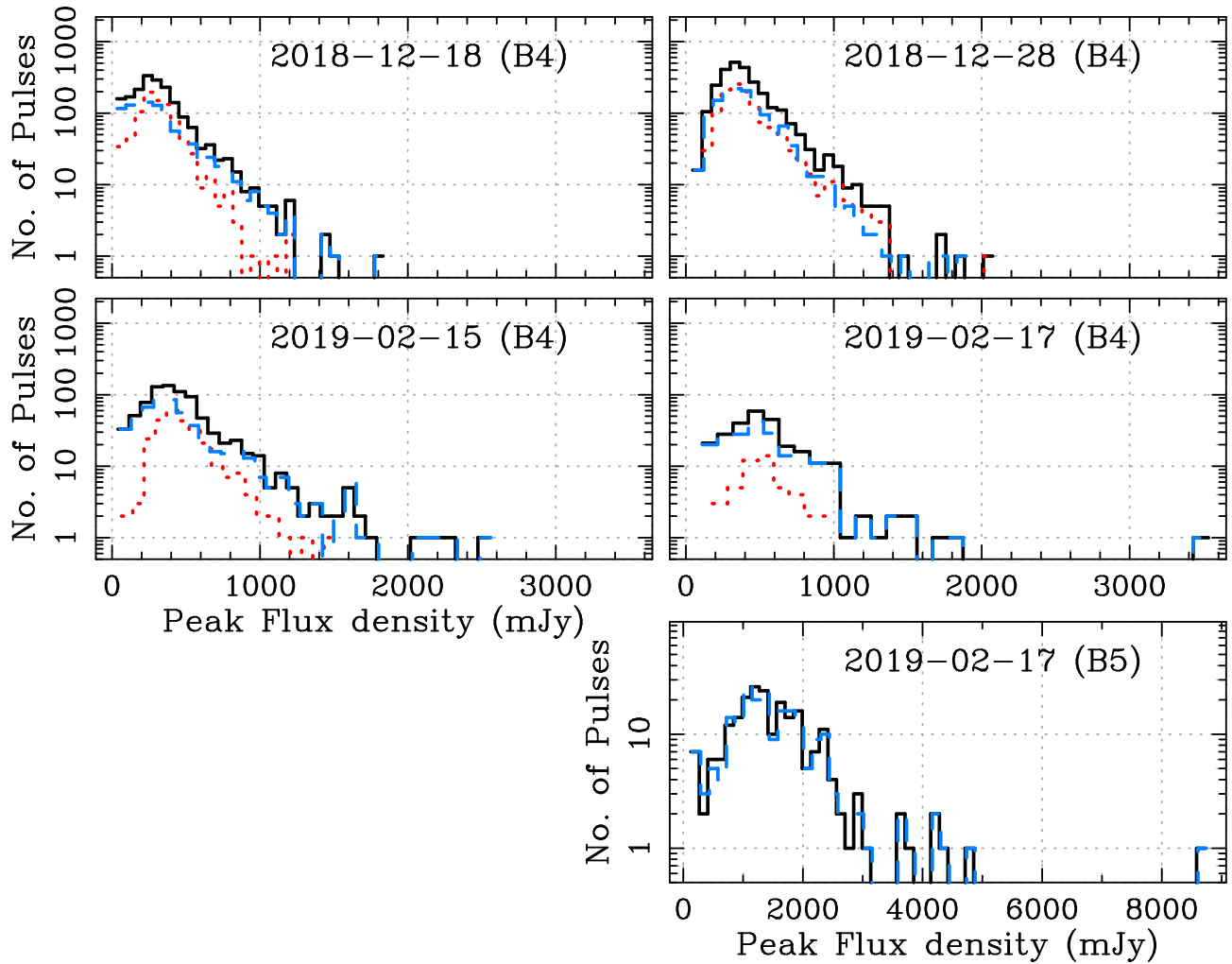}
\caption{The continuous black color histograms show the peak flux density
distributions of the bursty emission at the epochs and bands marked in the
respective panels. The red, dotted-line and the blue, dashed-line histograms
show the distributions for the bursts detected under the components M1 and M2,
respectively. Note that, for ease of comparison, the same flux density range
has been displayed for all the 650\,MHz (B4) observations. In the 1360\,MHz
(B5) observation, the number of bursts detected under M1 were too few to
compute a distribution.
  \label{fig-peakhist}}
\end{figure*}
\section{Analysis and Results} \label{sec-results}
%
\subsection{Flux density spectrum and evolution}\label{subsec-flux}
The period-averaged flux density was estimated by
using the `top-hat' equivalent width of the average intensity profile and
the corresponding signal-to-noise ratio (S/N) in the radiometer equation
\citep{handbook04}. We assumed 60\% aperture efficiency and receiver
temperatures as per the observatory specifications. The sky background
temperature (\tsky) towards the magnetar was estimated by scaling the
corresponding estimate at 408\,MHz \citep{Haslam82} to the center
frequency of our observations. At 650\,MHz, \tsky is estimated to be about
115\,K. The temporal evolution of the 650\,MHz flux density thus obtained
is shown in the top panel of Figure~\ref{fig-flux}.
\par
To obtain the spectrum, we used our December 18 (S1) observation
to compute average profiles in 4 different 50\,MHz wide sub-bands centered
at 575, 625, 675 and 725\,MHz. \tsky was estimated at each of these
frequencies as mentioned above.
The period-averaged flux densities at these sub-bands are shown in
Figure~\ref{fig-flux} (bottom panel), along with those in the frequency
range 768$-$3840\,MHz measured by \citet{Dai19} using the Parkes telescope
on the \emph{same day}. On December 21, we successfully detected
the magnetar in the frequency range 300$-$500\,MHz. Due to low S/N we
could not estimate the spectrum in this frequency range, however, we
have plotted the band-averaged flux density in Figure~\ref{fig-flux}.
\subsection{The spiky emission: Width and flux density distribution}
To understand the average properties of the bursty emission from the
magnetar, we used \texttt{single\_pulse\_search.py} from \textsl{PRESTO}
to first detect all the bright pulses above a detection threshold of
8$\sigma$. From our 650\,MHz observations, we detected a total of 1856,
2662, 818 and 261 bursts from sessions S1, S2, S3 and S4$_a$, respectively.
Additionally, we detected 219 bursts from the 1360\,MHz observation (S4$_b$).
The positions of these bursts, in terms of the magnetar's rotational phases,
could provide crucial clues to the underlying emission mechanism. As shown
in Figure~\ref{fig-sp} obtained from session S1, the number of detected
bursts roughly follow the total intensity profile shape. A similar trend
is noticed in the other observations too. Unlike the giant pulse emission
from several pulsars (e.g., B0531+21, B1937+21), these bursts are not confined
in narrow rotation phases inside or outside the average emission window.
\par
The spiky emission from the magnetar is known to be much narrower than
the individual components in the average profile. Our observations clearly
show that the bursts have a characteristic width of some 1$-$4\,ms at
650\,MHz (see Figure~\ref{fig-sp}). The characteristic width seems
to become narrower ($\lesssim$1\,ms) at 1360\,MHz. We note that the
observed pulse-width is essentially the intrinsic pulse-width convolved
with our sampling time. Peak of the 1360\,MHz distribution around
1.3\,ms may have been particularly affected by the coarse sampling time of
0.655\,ms. Due to this coarse resolution, pulses intrinsically as narrow as
0.7\,ms would also appear to be nearly 1\,ms wide. Hence, the actual
characteristic pulse-width at 1360\,MHz would be of the order of 1\,ms or
smaller. As discussed later in this section, the increase of the pulse-width
at lower frequencies is most likely caused by the propagation effects, and
not intrinsic.
\par
Following the convention used by \citet{Serylak09}, we mark three components
in the average profile: M1, M2 and M3 (see Figure~\ref{fig-sp}). Component~M3
is absent in session S1 (Figure~\ref{fig-sp}), but it appears in the other
sessions. However, even when M3 is visible, the number of spiky bursts detected
under this component are only 1$-$2\% of the total number of bursts. So, we
only consider the bursts detected under the components M1 and M2.
\begin{figure*}
\centering
\subfigure{\includegraphics[width=0.34\textwidth,angle=0]{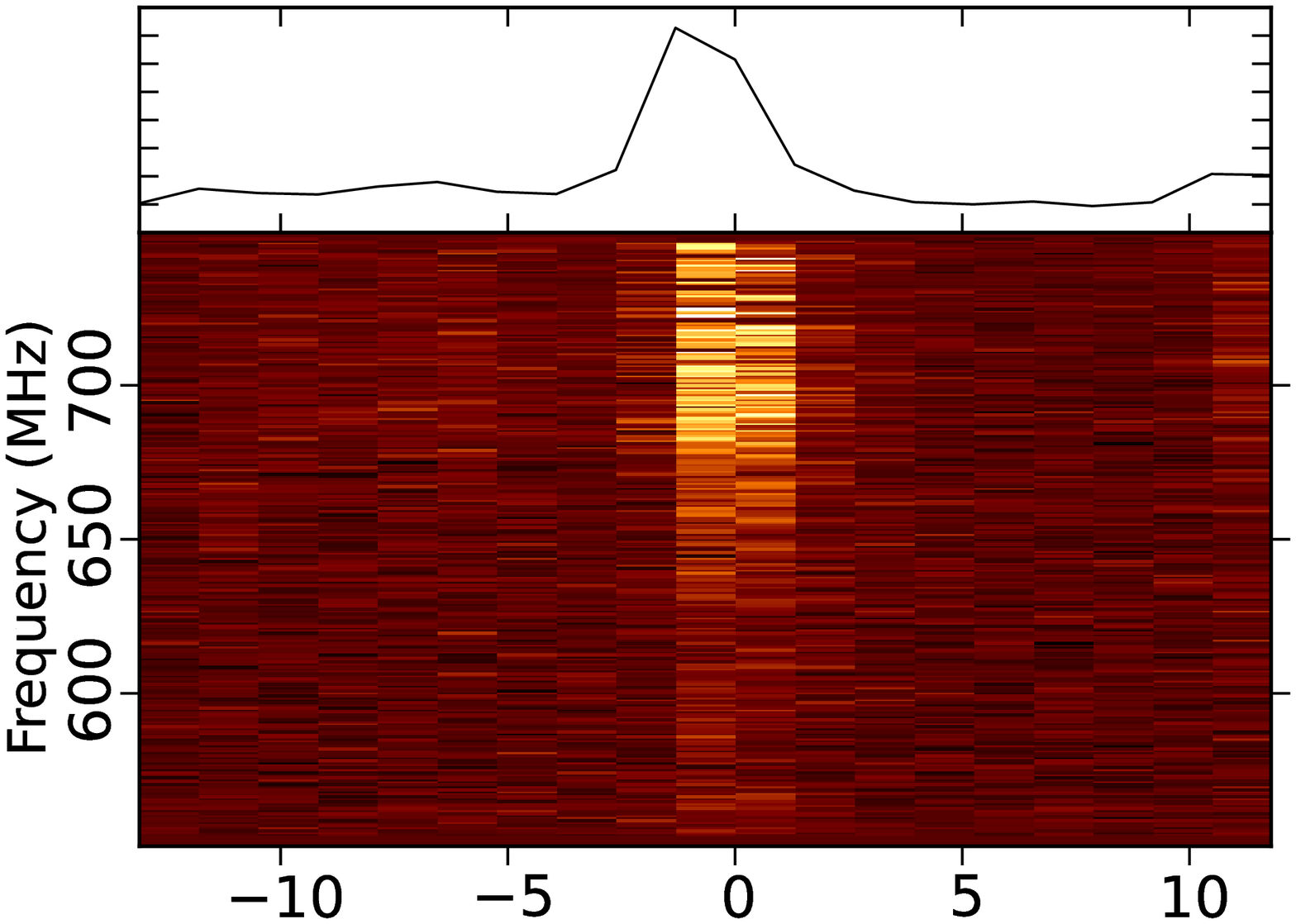}} \hspace*{-5mm}
\subfigure{\includegraphics[width=0.34\textwidth,angle=0]{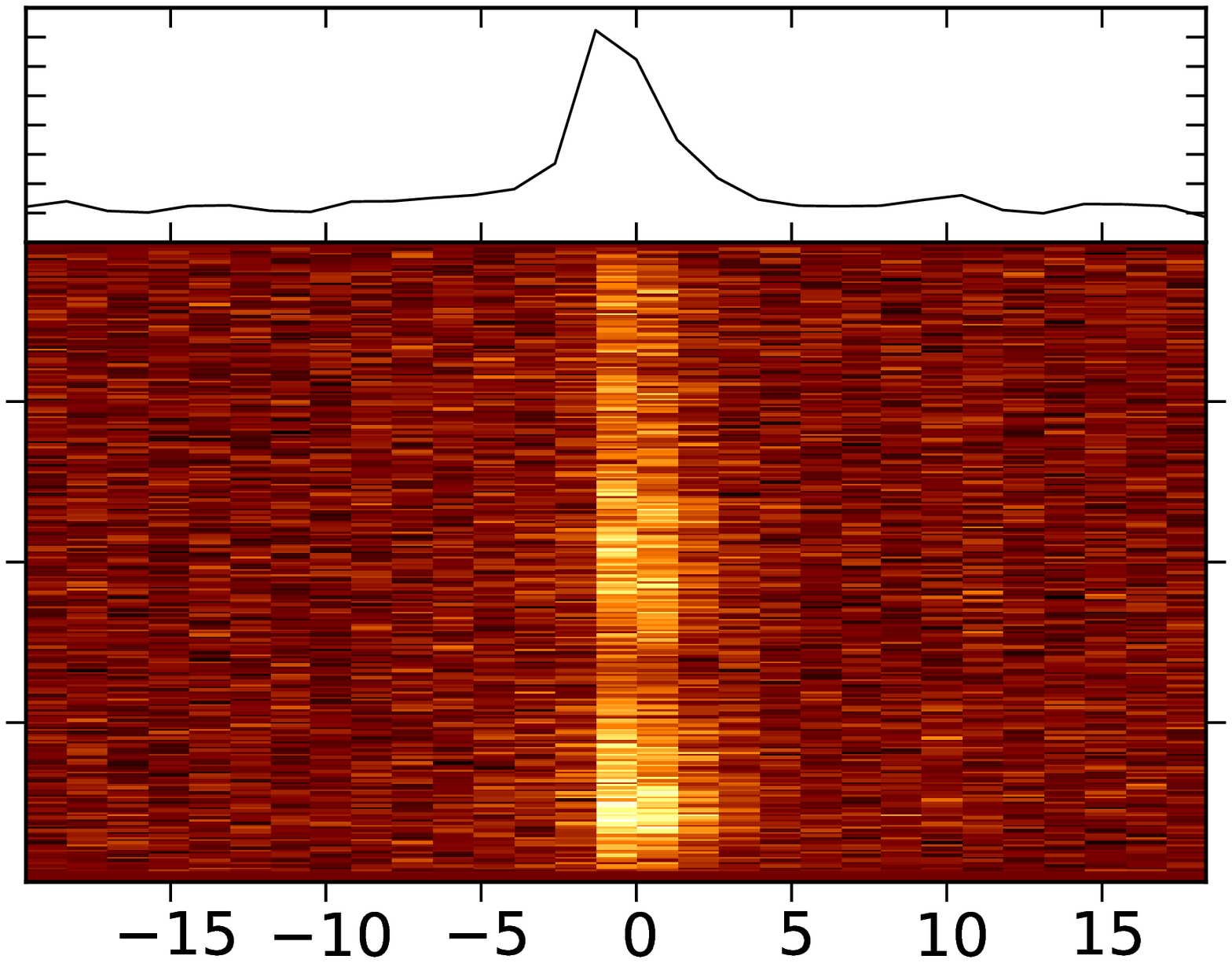}} \hspace*{-5mm}
\subfigure{\includegraphics[width=0.34\textwidth,angle=0]{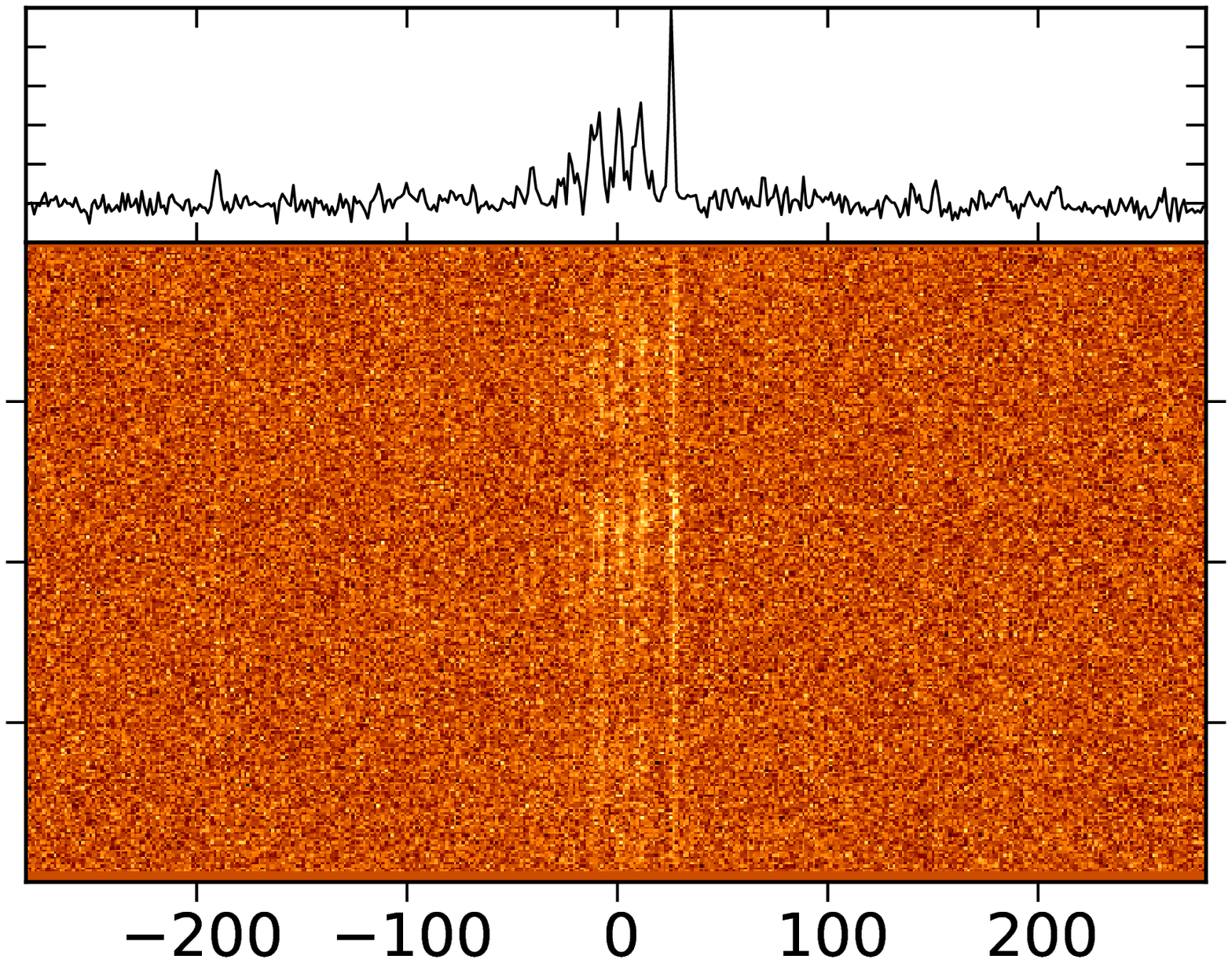}}
\subfigure{\includegraphics[width=0.34\textwidth,angle=0]{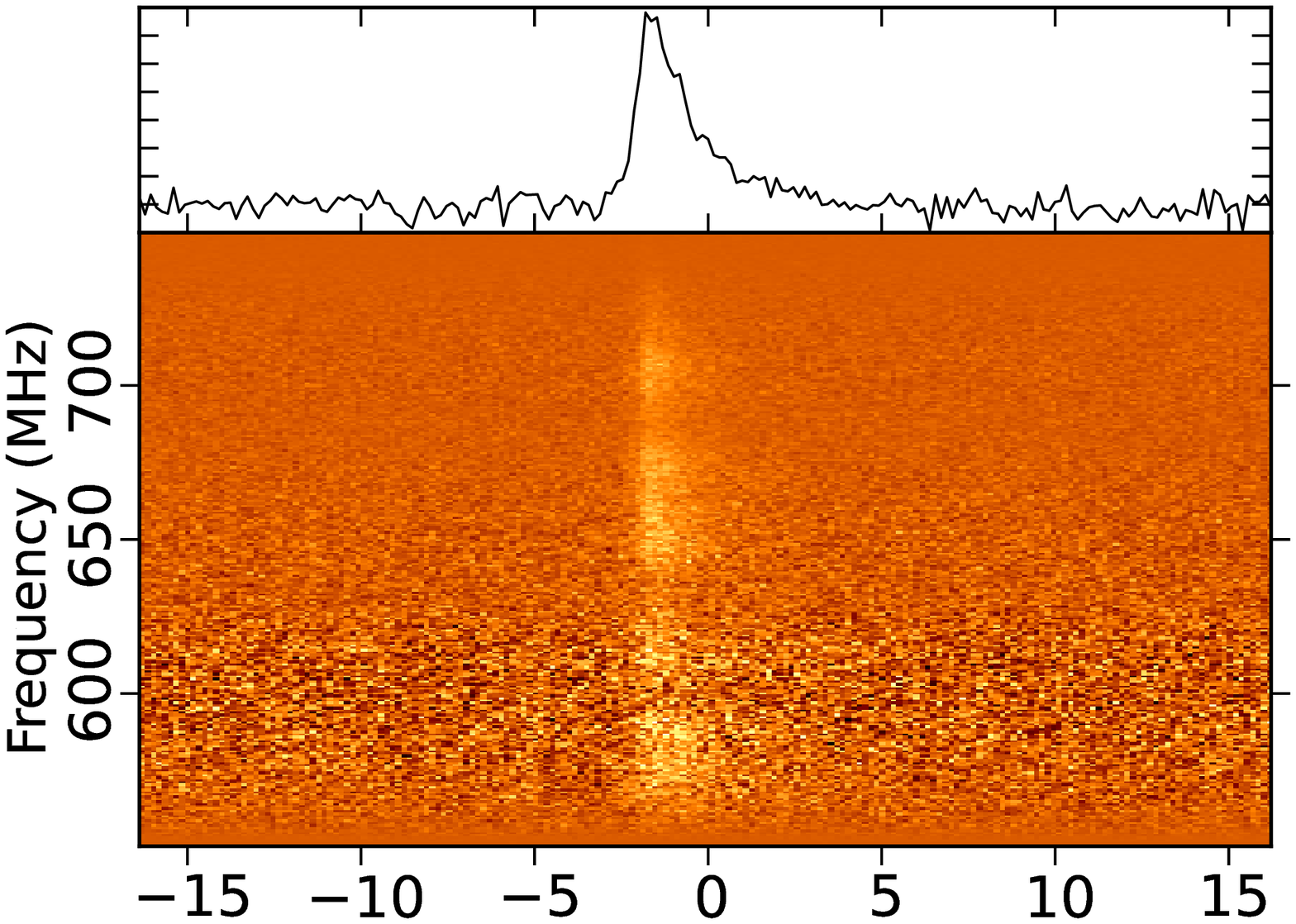}} \hspace*{-5mm}
\subfigure{\includegraphics[width=0.34\textwidth,angle=0]{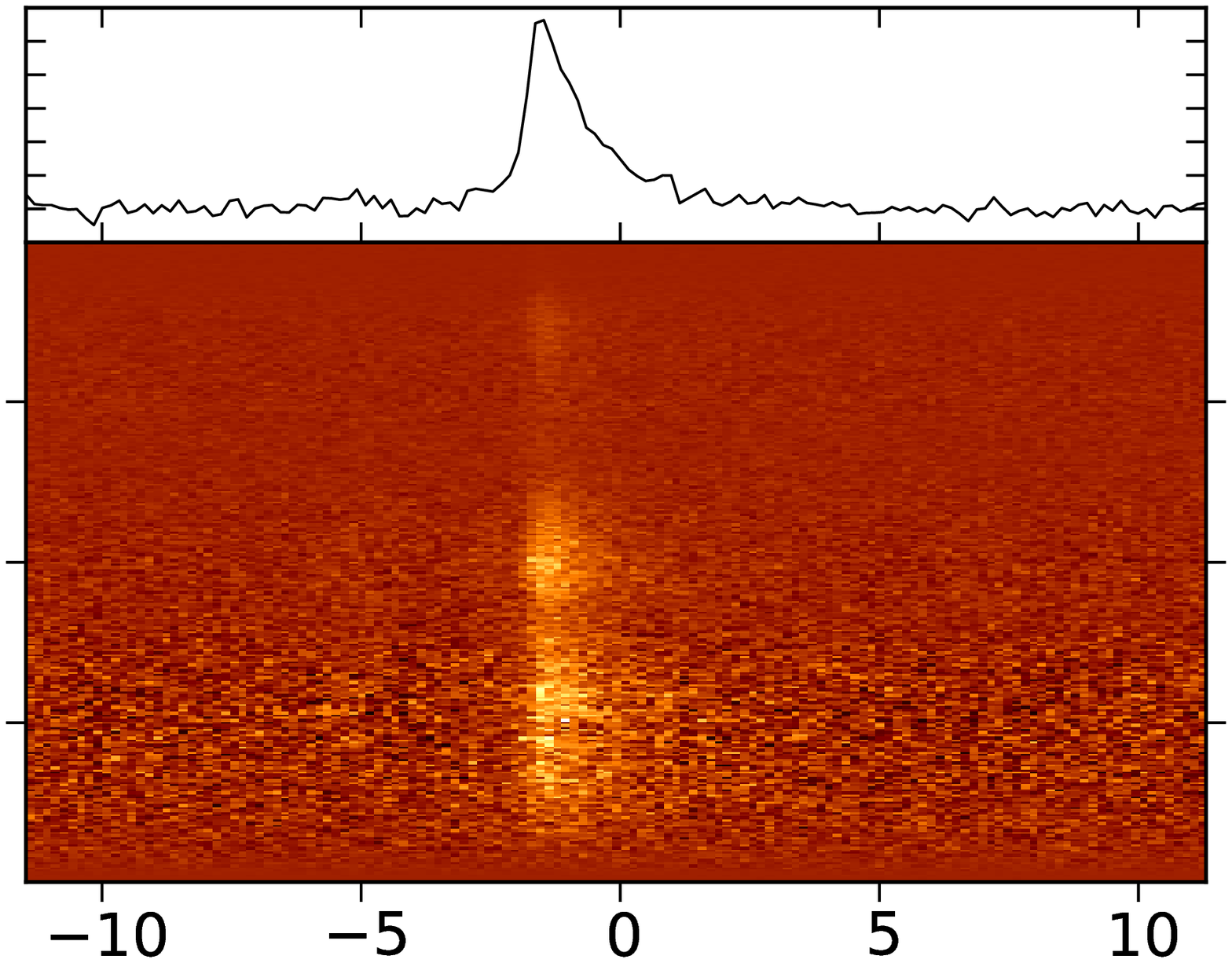}} \hspace*{-5mm}
\subfigure{\includegraphics[width=0.34\textwidth,angle=0]{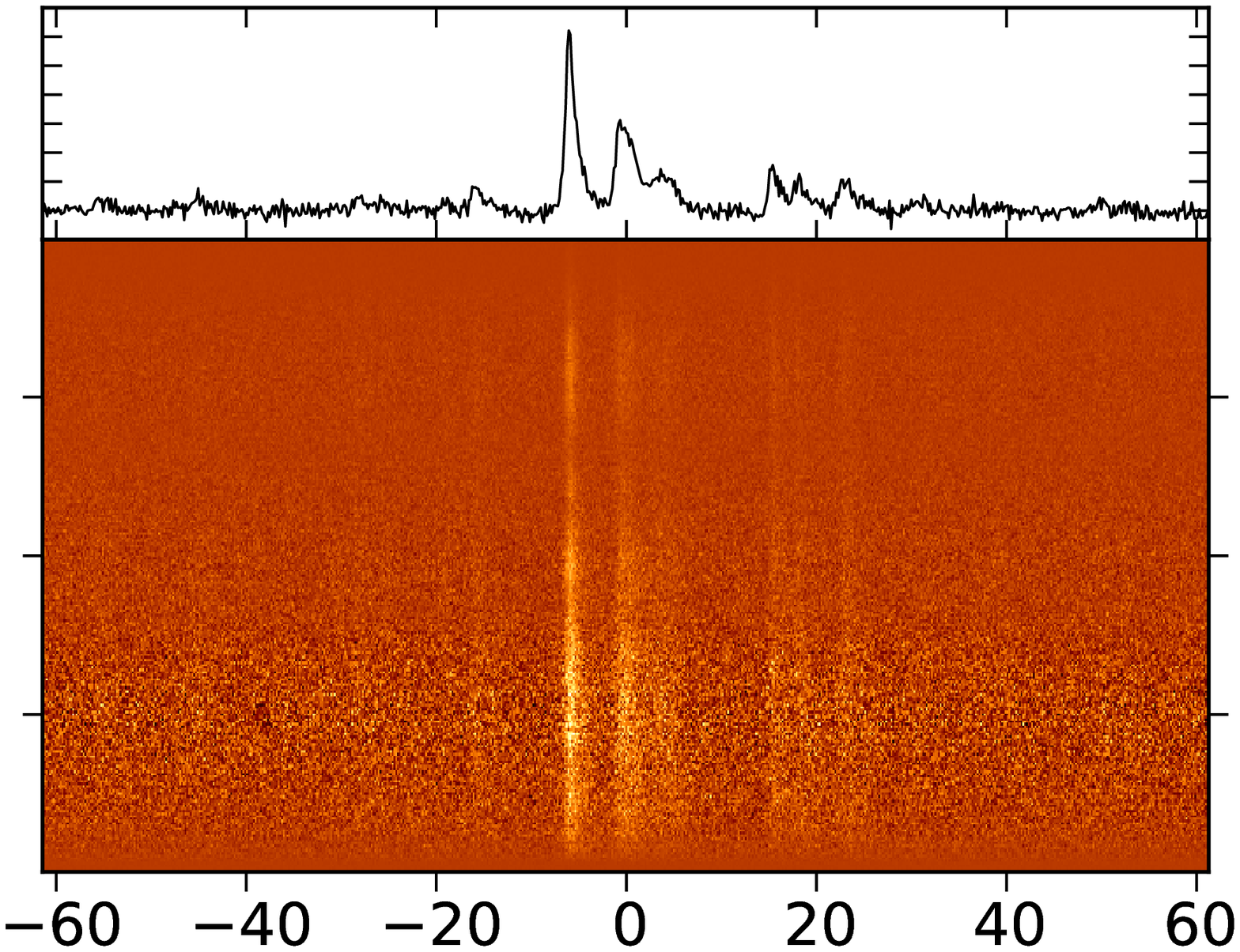}}
\subfigure{\includegraphics[width=0.34\textwidth,angle=0]{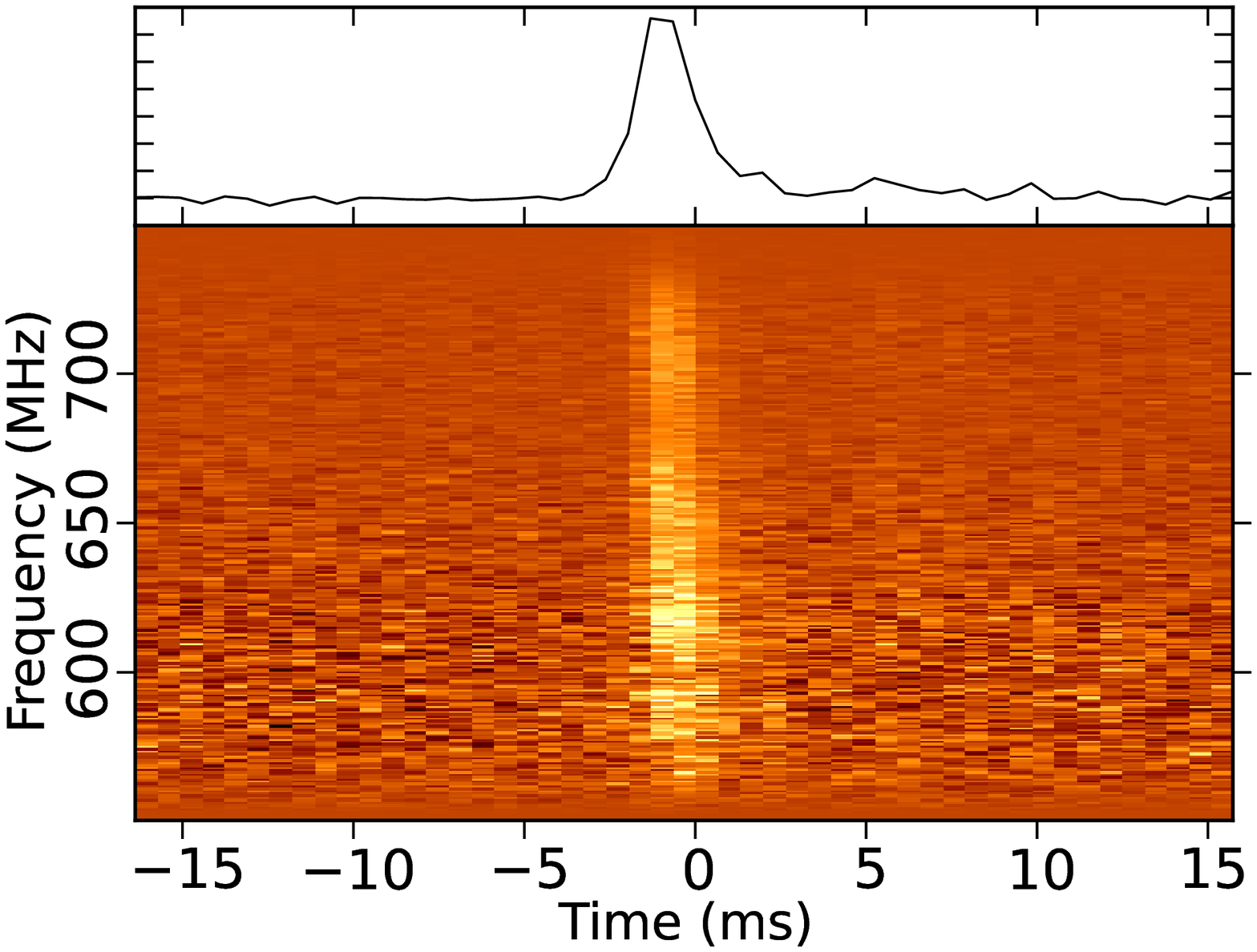}} \hspace*{-5mm}
\subfigure{\includegraphics[width=0.34\textwidth,angle=0]{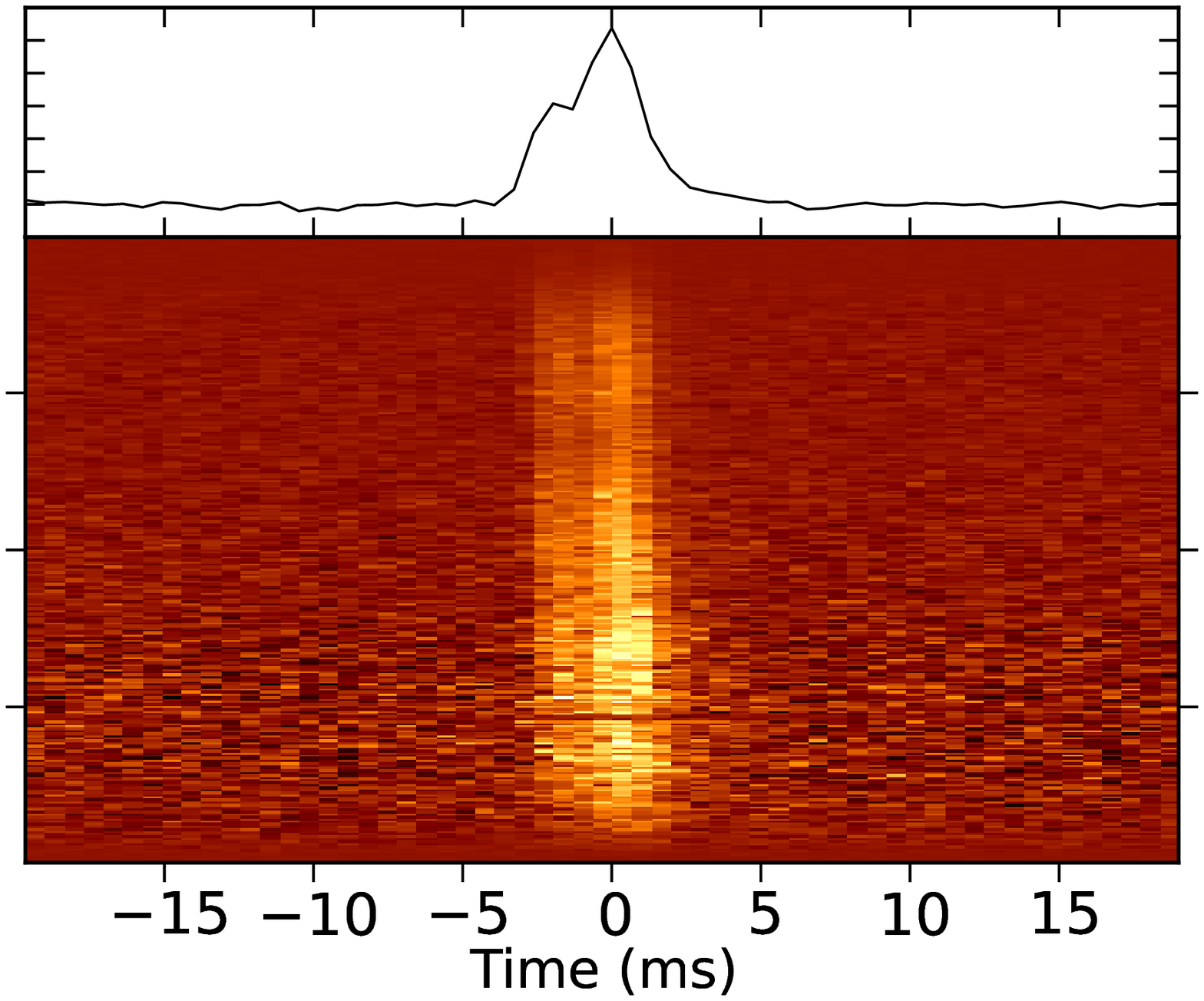}} \hspace*{-5mm}
\subfigure{\includegraphics[width=0.34\textwidth,angle=0]{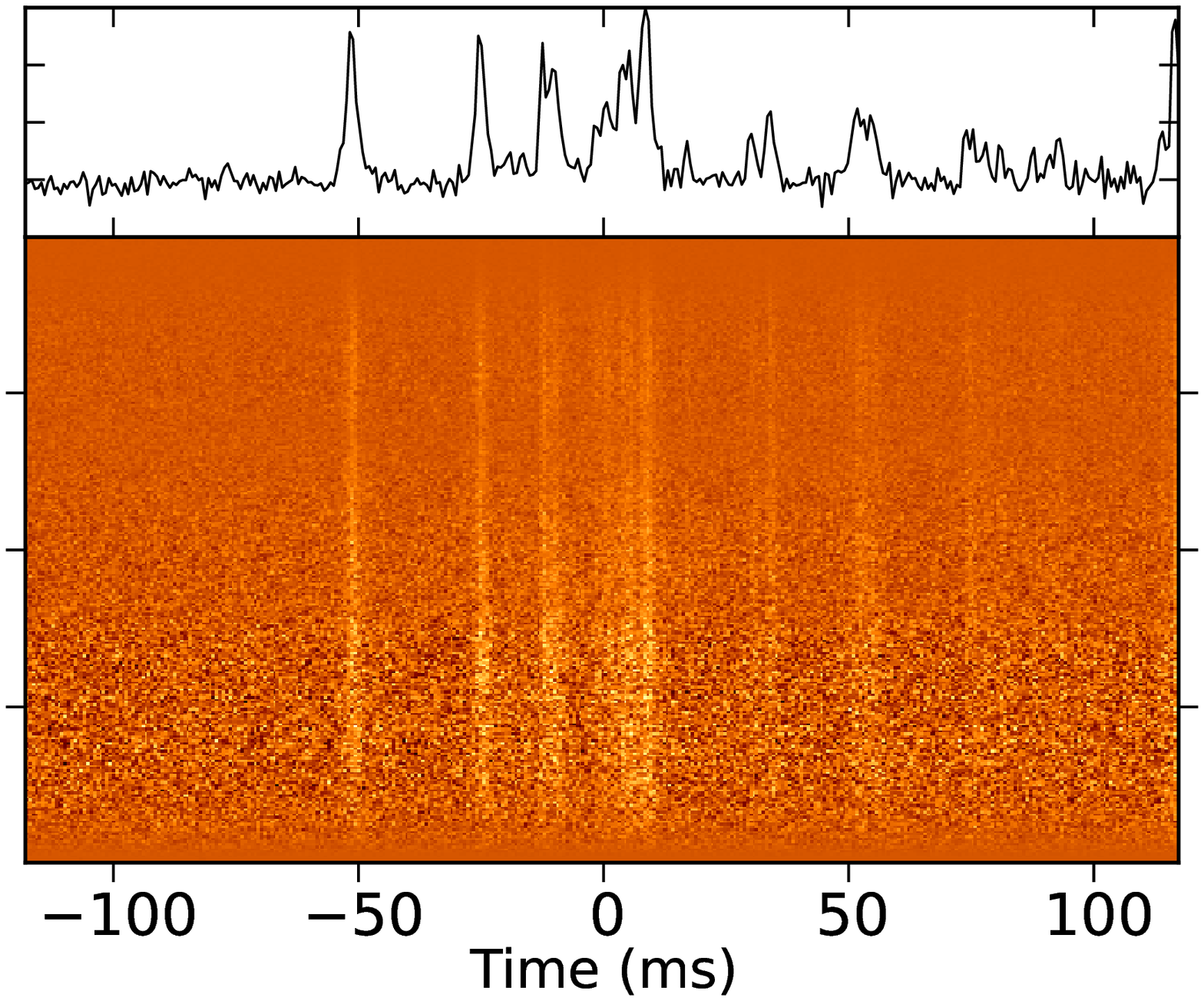}}
\caption{Spectro-temporal properties of a sample of bursts at 650\,MHz are
shown. The bursts in the first, second, and third rows are from the sessions
S1, S2, and S3, respectively.
\label{fig-spsample}}
\end{figure*}
\begin{figure}
\centering
\includegraphics[width=0.68\textwidth,angle=-90]{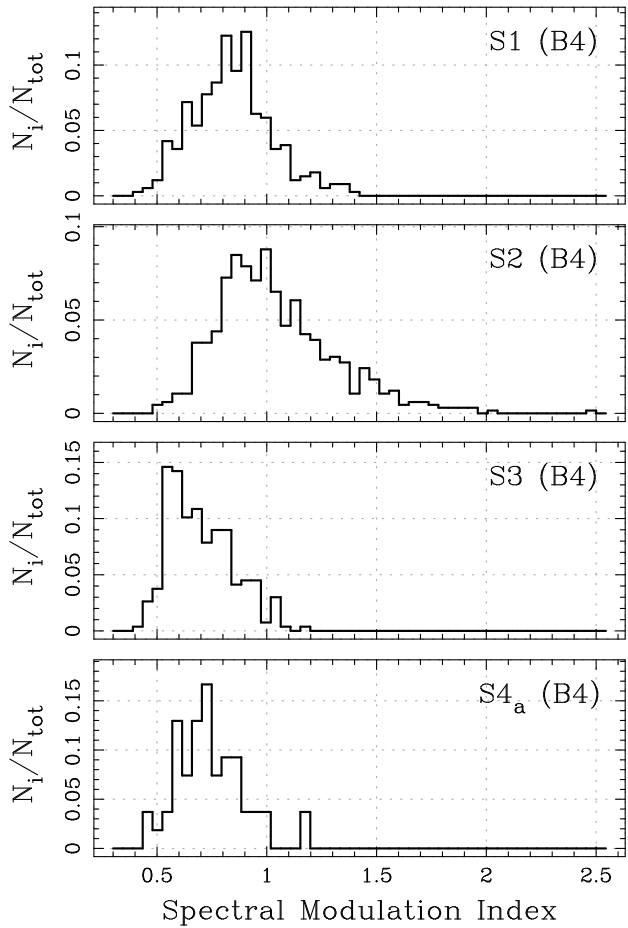}
\caption{The individual panels show the distribution of the spectral
modulation indices obtained for the peak-spectra of the bursts
from the 550$-$750\,MHz sessions marked in the respective panels.
\label{fig-mod}}
\end{figure}
%
\par
To determine the peak flux density ($S_p$) of the individual pulses \citep{CM03},
we use the pulse-width and the peak~S/N of the pulse corresponding to a
smoothing optimum for its observed width, in the modified radiometer
equation as described in \citet{MA14}. The $S_p$ distributions of the
bursts under the two components, combined as well as component-separated,
in individual observations are presented
in Figure~\ref{fig-peakhist}. We model the tails of the distributions
using power-low statistics of the form: $N(S_p)\propto S_p^{\alpha}$, where
$N$ is the number of pulses, and
$\alpha$ is the power-law index. Furthermore, we used $S_p=500$\,mJy as
a uniform lower-cutoff to fit only the tails of the distributions. The
fitted values of $\alpha$ for the overall burst distributions are presented
in Table~\ref{tab1}. The distributions in the last two observations appear
to be slightly flattened when compared to those from the first two observations
which were much closer to the start of the outburst. Except for the session~S2,
the 650\,MHz bursts-distributions under M1 have significantly steeper tails
than those of under M2 (Figure~\ref{fig-peakhist}). As evident by a single
point at the far right tail of the distributions from sessions S4$_a$ and S4$_b$,
a narrow, very bright pulse with peak flux densities of about 3.5\,Jy
and 8.6\,Jy was detected at 650 and 1360\,MHz, respectively.
\subsection{Spectro-temporal characteristics of the bursts}
The bursts from the magnetar show a variety of temporal and spectral
characteristics and phenomenology.
A representative set of bursts from our 650\,MHz observations are presented
in Figure~\ref{fig-spsample}. The first two panels in each of the rows show
some of the narrowest bursts and their spectral structures in the respective
sessions. Most of these bursts exhibit an exponential tail,
indicating scatter broadening in the intervening propagation medium. To
determine the intrinsic width and scatter broadening timescale (\tausc), we
modelled the first two bursts from session~S2 shown in Figure~\ref{fig-spsample}
as a Gaussian convolved with a one-sided exponential function \citep{KK19}.
Our model fits suggest the intrinsic widths of the two bursts to be
$0.69\pm0.05$\,ms and $0.53\pm0.04$\,ms, and the corresponding \tausc
to be $1.30\pm0.06$\,ms and $1.05\pm0.05$\,ms, respectively. So, the
apparent characteristic pulse-width of a few ms at 650\,MHz is predominantly
due to scatter broadening. The above intrinsic pulse-widths at 650\,MHz
are consistent with the characteristic pulse-width of less than 1\,ms at
1360\,MHz, as discussed earlier.
\par
The bursts also show significant spectral variations (Figure~\ref{fig-spsample}).
Interstellar scintillation could induce such spectral structures. However,
using the above measured \tausc, the scintillation bandwidth is estimated
to be less than a kHz. Both the popular electron density models,
NE2001 \citep{CL02} and YMW16 \citep{YMW17},
also suggest similar estimates. Hence, the observed spectral structures
of several tens of MHz are intrinsic to the source. The structures are more
prominently noticeable closer to the outburst onset (i.e., in sessions S1
and S2) and less so in the later
observations. The frequency bandwidths of these structures seem to have narrowed
down with time. The bursts in the later sessions appear to have more and more
uniform and featureless spectra. The number of bursts which show prominent
spectral variations such as those in the first row, also appear to be much
lesser in the later sessions.
To quantify the spectral variations, we compute spectral modulation
index, \mi, defined as $\misq=(\isqbar-\ibarsq)/\ibarsq$, where \ibar and
\isqbar are the first and second moments of the peak-intensity of a burst
as a function of frequency \citep{Spitler12}. For each of the sessions, we
selected all the bursts narrower than 3\,ms and with S/N$>$10. We then
computed \mi for the dedispersed spectra corresponding to the peaks of the
bursts after partially averaging in frequency to obtain 128 sub-bands across
the bandwidth. The obtained \mi are in the ranges 0.40$-$1.41, 0.49$-$2.46,
0.42$-$1.20 and 0.46$-$1.19, using 335, 660, 267 and 54 bursts from the
sessions S1, S2, S3 and S4$_a$, respectively. The distributions of the
obtained \mi are shown in Figure~\ref{fig-mod}. Note that the lowest values
of \mi in the above ranges correspond to the bursts where the spectral power
is near uniformly distributed across the frequency band, and, as expected,
these are indeed quite close to each other (0.40$-$0.49). A higher value
of \mi implies the spectral power to be localized in smaller sub-bands. As
apparent from Figure~\ref{fig-mod}, the sessions S1 and S2 have many more
bursts with relatively higher \mi, and hence, with more spectral variations
than the last two sessions. Moreover, in the
last two sessions (S3 and S4$_a$), peaks of the distributions have shifted
very close to the lowest observed \mi values indicating that majority of the
bursts in these sessions have close to uniform and featureless spectra.
\par
As apparent from the last panels in each of the rows in
Figure~\ref{fig-spsample}, many times the bursts tend to occur in succession.
Sometimes even a quasi-periodic occurrence is noticeable, with separations
in some 5$-$25\,ms range. A more quantitative characterization of the
quasi-periodicity as well as the above spectral structures (e.g., measuring
the structure bandwidths, and their evolution with time and frequency) will
be reported elsewhere.
\section{Discussion} \label{sec-discuss}
After the previous outburst in 2003, the radio spectrum of \magn was found
to be near flat or mildly steep, with a flux density power-law index
$-0.5\lesssim\alpha\lesssim0$ \citep{Lazaridis08,Camilo07c}. However,
\citet{Dai19} suggest the spectrum to be slightly harder after
the onset of the current outburst. Their measurements on December~18
suggest {$\alpha$=+0.8$\pm$0.1} in the frequency range 768$-$1800\,MHz.
If we include our flux density measurements below 750\,MHz on the same
day, we obtain {$\alpha$=+1.2$\pm$0.1}. This indicates that the spectrum
is perhaps even harder at lower frequencies.
\par
The period-averaged flux density has decreased rapidly since the onset
of the outburst. Similar to the previous outburst \citep{Camilo16}, the
650\,MHz flux density decreased by a factor of about 5 or more in the
first 20$-$30 days. The flux density at 1.52\,GHz also shows a similar
trend \citep{Levin19}. \citet{Camilo16} also showed an anti-correlation
between the flux density and the spin frequency derivative during the
previous outburst.
However, the spin frequency derivative and 1.52\,GHz flux
density estimates from
\citet[][see the upper two panels in their figure~6]{Levin19}
rather show a
correlated behavior of the two parameters in the current outburst,
making any possible physical link between the two further unclear.
\subsection{Spiky emission: giant pulses or giant micropulses?}
A working definition of giant pulses is that the flux density or
pulse-energy of a single pulse averaged over the spin period is
more than 10 times the corresponding mean quantity. The mean
flux density of the bursts from the magnetar do not exceed the
above conventional threshold. Nevertheless, the peak flux densities
of the bursts are very large in absolute terms. For example, the
brightest pulses in session S3 and S4 are about 2.5 and 3.5\,Jy,
which is 40$-$60 times the mean $S_p$ of the average profiles.
These properties are rather reminiscent of the giant micropulses
\citep{Johnston01} discovered from the Vela pulsar (B0833$-$45).
\par
Much like the giant pulses, the giant micropulses add extended
power-law tails to the single pulse flux density distributions
\citep{KJvS02}. The peak flux densities are often averaged over
the period before making the histograms \citep[e.g., see][]{Karuppusamy10}.
We note that the distributions of the absolute (Figure~\ref{fig-peakhist})
as well as those of the period-averaged (not shown but analyzed
separately) peak flux densities exhibit tails which are very well
fit with a power-law.
\par
\citet{KJvS02} reported a characteristic relationship between
widths of micropulses and the spin period. Using their equation~1, 
the expected micropulse width for \magn would be 2.8$-$5.6\,ms.
While this width is consistent with the characteristic width of the
bursts we measure at 650\,MHz, we note that the intrinsic widths
of the narrowest pulses are only about 0.5$-$0.7\,ms, i.e., smaller
by a factor of 4$-$5. However, the widths of the narrowest giant
micropulses from the Vela pulsar are in fact also smaller than
those of the average micropulses by a similar factor \citet{KJvS02}.
In any case, given the scatter in the data used to derive the
relationship between micropulse width and the spin period, the
characteristic width of the magnetar's bursts is consistent with
the expected micropulse width.
The micropulses often exhibit an associated quasi-periodicity
in their occurrences. While a detailed analysis of any underlying
quasi-periodicity is under progress, Figure~\ref{fig-spsample}
shows some examples of a possible underlying quasi-periodicity
in the magnetar's bursts. 
\par
The giant micropulses from Vela pulsar and the classical giant
pulses from a handful of pulsars occur in narrow pulse phase
ranges. However, the giant pulses from the Crab pulsar appear
in significant parts of the pulse window. The bursts from the
magnetar also do not have any favorable phase-ranges, and their
occurrence rate roughly follow the integrated profile shape.
\subsection{Any links with FRBs?}
While most FRBs are one-off events, a couple of these are known to repeat
\citep[\rfrb and FRB~180814.J0422+73;][]{Spitler16,CHIME19a}. Recently,
\citet{Hessels19} used high time resolution observations to show complex
time$-$frequency structures in the \rfrb bursts. Particularly, they showed
that many of the bursts at 1.4\,GHz show $\sim$250\,MHz wide frequency bands
which cannot be explained by scintillation in the ISM. Similar structures
have been seen in the second repeating FRB as well \citep{CHIME19a}.
Except for the
high-frequency interpulse giant-pulses from the Crab pulsar \citep{Hankins16}
and in some faint emission components of PSR~1745$-$2900 \citep{Pearlman18},
such frequency structures have not been seen from any known pulsar or
magnetar. In Figure~\ref{fig-spsample}, we show that the spiky emission
from \magn also exhibits frequency structures which cannot
be caused by the ISM scintillation. \citet{Hessels19} also showed that
the banded structures in many of the \rfrb bursts show a frequency drift,
and the drift rates possibly decrease at lower frequencies.
The relatively coarse time-resolution of our observations
and the smearing caused by the scattering at our observing frequency does
not allow adequate probes of
any underlying drifts at timescales shorter than a few ms.
However, the frequency drifts of FRB~180814.J0422+73
are observed at much longer timescales of some 20$-$60\,ms in the frequency
range 400$-$800\,MHz. We do not notice such long timescale drifts in the
magnetar's bursts.
In any case, high time resolution probes of the spiky emission from the
magnetar at adequately high frequencies will clarify if the frequency
structures indeed share some similarities with those from the
repeating FRBs, for which magnetar origin is one of the popular models.
\par
We note that the 1.36\,GHz peak flux density of the brightest burst in our
sample is about 9\,Jy. This is about an order of magnitude larger than the
peak flux density of the bright bursts from \rfrb at similar frequencies.
However, \rfrb is nearly 2.8$\times$10$^5$ times more distant, implying a
$\sim$10$^{11}$ times more luminosity, for similar emission solid angles.
Nevertheless, the fact that the magnetar \magn is only the third object after
the repeating FRBs and the Crab pulsar which is found to exhibit frequency
structures in its bursts so prominently, might provide a phenomenological link between the
underlying emission mechanisms.
\section{Conclusions} \label{sec-con}
To conclude, we have presented the spiky emission properties of the magnetar
XTE~\magn
as well as its flux density evolution and low-frequency spectrum in the early
phases of the recent outburst (December 2018). We have shown that the bursts
from the magnetar exhibit frequency structures that cannot be explained by
interstellar scintillation. The spectral structures are easily noticeable in
the early phases of the outburst, and seem to fade away in the later phases.
The energetics of the spiky bursts show similarities with those of the
giant micropulses, which may indicate a link between the corresponding
underlying emission mechanisms. More detailed study of the spiky emission,
including characterization of the frequency structures and their temporal
evolution is under progress, and will be a subject of future publication.
\acknowledgments
We thank the anonymous referee for constructive comments which helped in
improving the manuscript.
YM thanks M. A. Krishnakumar for help with a software module which was useful
in estimating the scatter broadening timescale. YM acknowledges use of the
funding from the European Research Council under the European Union's Seventh
Framework Programme (FP/2007-2013)/ERC Grant Agreement no.~617199.
MPS acknowledges support from NSF RII Track I award number OIA$-$1458952.
MPS is a member of the NANOGrav Physics Frontiers Center which is supported
by NSF award 1430284.
We thank the GMRT observatory for allocation of director's discretionary time
for this project.
We thank the staff of the GMRT who have made these observations possible. The
GMRT is run by the National Centre for Radio Astrophysics of the Tata
Institute of Fundamental Research.

\vspace{5mm}
\software{PRESTO \citep{RansomThesis}}
\facility{GMRT(GWB)}

\end{document}